\documentstyle[preprint,prc,aps,epsfig]{revtex}
\begin{document}

\title{Parity splitting in the alternating parity bands of some actinide
       nuclei.}
\author{R. V.~Jolos$^{1,2}$, and P.~von Brentano$^1$}
\address{$^1$ Institut f\"ur Kernphysik, Universit\"at zu K\"oln, 50937 
K\"oln,Germany}

\address{$^2$ Bogoliubov Theoretical Laboratory, Joint Institute for 
Nuclear
Research, 141980 Dubna, Russia}

\date{\today}
\tighten
\maketitle

\begin{abstract}
An angular momentum dependence of the parity splitting in the 
alternating
parity bands of nuclei with strong octupole correlations is considered
basing on the model of the octupole motion in a one--dimensional 
potential
well conserving axial symmetry. A sign reversal of the parity splitting 
at
higher values of the angular momentum is interpreted as a result of the
Coriolis coupling to $K$=1 octupole mode. New experimental data on the 

spectra of the alternating parity octupole bands are used.
\end{abstract}

\pacs{21.10.Re, 21.60.Ev} 

\section{Introduction}

The presence of strong octupole correlations in nuclei is reflected
in a significant lowering of the excitation energies of the negative 
parity
states. This phenomena was observed in different nuclei 
\cite{Butler,Ahmad}.
Among them in $^{222,224,226}$Ra and $^{224,226}$Th isotopes which
are considered as rotating octupole--deformed systems. In these nuclei
low--lying negative parity states form the alternating parity bands
together with the low--lying positive--parity states connected via
strong E1 transitions. It is known that
such rotational bands with $I^{\pi}=3D0^+,1^-,2^+$,... exist in
molecules. However, in nuclei in contrast to molecules these bands are
disturbed: the negative--parity states are shifted up in energy with
respect to the positive--parity states. Especially interesting is the
angular momentum dependence of this parity splitting. At the beginning 
of
the rotational band the positive-- and negative--parity states are
split in energy by several hundreds of $KeV$. Even in octupole--deformed 
nuclei the parity splitting is 200--300 $KeV$. At higher values of the 
angular
momentum the parity splitting decreases and the energies of both
positive-- and negative--parity states start to have approximately
the rotational angular momentum dependence appropriate to a $K$=0 
band.
Thus, a unified
alternating parity band is forming. However, not perfectly. Namely, at 
some
value of the angular momentum the parity splitting changes sign and the
negative--parity states are lowered slightly below the positive--parity
states. This effect is, however, 5--10 times smaller than the positive 
parity
splitting observed at the beginning of the rotational band.
With further increase
of the angular momentum the value of this negative
parity splitting decreases.
Thus, there are two different effects connected with parity
splitting which manifest in the different ranges of the angular
momentum scale. The large positive parity splitting at low angular
momentum decreases with
angular momentum where as the negative parity splitting
which is 5--10 times smaller than the positive parity splitting
appears at higher values of the angular momentum.
To see both effects it is important to have the experimental
data on both positive-- and negative--parity rotational states up to
sufficiently high values of the angular momentum $I$. Beautiful data of
this kind have been obtained recently by the Liverpool group
\cite{Cocks}. The positive parity splitting was quantitatively
interpreted in the framework of the model developed by us in
\cite{Jolos1,Jolos2,Jolos3} for the angular momentum range available at
that time. In the present paper we shall apply the model to the 
description
of the new data of the Liverpool group.
We will show also how to generalize the model to
incorporate the effect of the sign reversal of the parity splitting. As
was discussed in \cite{Jolos1,Jolos2,Jolos3} the behaviour of the
parity splitting from low to moderate values of $I$, where the effect is
largest, can be explained in the framework of the one--dimensional
model of the octupole motion conserving axial symmetry. We will show
below that the smaller effect of the negative parity
splitting can be explained by Coriolis coupling of the negative
parity states of the $K$=0 band to the octupole modes
with $K\neq$0.

The model developed below is a phenomenological model. In principle,
parity splitting can be obtained naturally in the framework of the
microscopic models with parity projection. For example, see the work
by the Madrid group \cite{Egido}, which employ selfconsistent mean field
approach with parity projection.

\section{Double minimum potential model}

To explain the angular momentum dependence of the parity splitting at
low $I$ the following model has been suggested in 
\cite{Jolos1,Jolos2,Jolos3}.
It was assumed that at low to moderate angular momenta
the parity splitting is
connected with the octupole motion in a one--dimensional potential well
depending on the collective variable $a_{30}$ which describes the 
axially
symmetric octupole deformation. In nuclei with sufficiently strong
octupole correlations or in octupole--deformed nuclei this potential has
two minima which are symmetrically located at positive and negative
values of $a_{30}$. The two minima are separated by the barrier and the
transition frequency corresponds to the parity splitting
in the alternating parity bands. The height of the barrier
depends on the angular
momentum via a rotational energy term
$\frac{\hbar ^2}{2\Im}I(I+1)$.
Due to dependence of the moment of
inertia on the octupole deformation parameter $a_{30}$ 
\cite{Nazarewicz1,Nazarewicz2,Nazarewicz3,Nazarewicz4,Egido}
the effective barrier
height between two minima increases with angular momentum $I$ 
proportional
to $I(I+1)$ thus decreasing
the probability of the barrier penetration. As a result the octupole
deformation stabilizes and the parity splitting goes to zero.
For physically reasonable shapes of the octupole
potential it was shown that the parity splitting can be quite well
described as an exponentially decreasing function of the barrier height.
This result has been exploited to parameterize the angular momentum
dependence of the parity splitting  $\Delta E(I)$ (this quantity was
introduced at first in \cite{Nazarewicz2})
in a simple fashion
\begin{eqnarray}
\widetilde{\Delta E(I)}=\widetilde{\Delta E(0)} exp(-\frac{I(I+1)}{I_0 
(I_0 +1)})
\label{deltae_eq}
\end{eqnarray}
(In the previous paper \cite{Jolos3} the notation $J_0$ has been used, 
however,
instead of $I_0$)
This formula has been suggested in \cite{Jolos2}. It contains two
parameters and describes a smooth decrease of the parity splitting 
without
changing of the sign of $\widetilde{\Delta E(I)}$.

\section{Coriolis coupling to the intrinsic state with $K^{\pi}=1^-$}

The importance of the bands coupling effect for the description of the
rotational alternating parity band was discussed in \cite{Nazarewicz3}.
In this section we consider phenomenologically a possible effect of
the Coriolis coupling of the $K^{\pi}=0^-$ and  $K^{\pi}=1^-$
states. To do this we must have a model for the intrinsic state
wave functions. A
suitable formalism is the coherent state method \cite{Vitturi,Raduta}.
To use it we introduce at first creation and annihilation operators
of the intrinsic phonons
\begin{eqnarray}
a_{30}\sim (b^+_{30} + b_{30} )
\label{phonon_eq}
\end{eqnarray}
The intrinsic wave functions of the $K$=0 states with even angular
momentum (positive parity) and odd angular momentum (negative parity)
have different parities with respect to reflection.
We will take the following expressions for them
\begin{eqnarray}
|\pi =(-1)^I , K=0>=\frac{1}{\sqrt{2(exp(\lambda ^2) + (-1)^I 
exp(-\lambda ^2))}}\nonumber\\ \times\left (exp(\lambda b^+_{30}) 
+(-1)^I exp(-\lambda b^+_{30})\right )|0>,
\label{coherent1_eq}
\end{eqnarray}
where $b_{30} |0>=0$.
In (\ref{coherent1_eq}) the parameter $\lambda$ is proportional to the 
octupole
deformation parameter. When $\lambda$ goes to zero we get
\begin{eqnarray}
|\pi =+1, K=0>\hspace{0.5cm} \rightarrow 
\hspace{0.5cm}|0>,\nonumber\\
|\pi =-1, K=0>\hspace{0.5cm} \rightarrow \hspace{0.5cm} b^+_{30}|0>
\label{states1_eq}
\end{eqnarray}
Thus, we get the limit of the harmonic octupole vibrations. It is seen 
that
the states (\ref{coherent1_eq}) with even and odd $I$ have a different
parity with respect to the $a_{30}\rightarrow -a_{30}$ reflection, i.e.
with respect to the $b^+_{30}\rightarrow -b^+_{30}$ transformation. It 
is
easy to check and it is seen already from the limit (\ref{states1_eq})
that odd angular momenta states have a larger average number
of the $K=0$ octupole
phonons than even angular momenta states.
This circumstance is important for the
consideration below. From this it follows that the Coriolis matrix
elements are larger for odd angular momenta  states.
For very large $\lambda$ (strong
octupole deformation) this difference goes to zero.

For the intrinsic states with $K=1$ we use by analogy to 
(\ref{coherent1_eq})
the following expression
\begin{eqnarray}
|\pi =(-1)^I , K=1>=\frac{1}{\sqrt{2(exp(\lambda ^2) - (-1)^I 
exp(-\lambda ^2))}}\nonumber\\ \times\left (exp(\lambda b^+_{30}) 
-(-1)^I exp(-\lambda b^+_{30})\right )b^+_{31}|0>,
\label{coherent2_eq}
\end{eqnarray}
The Coriolis interaction has a standard form
\begin{eqnarray}
H_{Coriolis}=-\frac{\hbar ^2}{2\Im}\left( J_+ I_- + J_- I_+ \right ),
\label{coriolis1_eq}
\end{eqnarray}
where
\begin{eqnarray}
J_+=\sum_K <K+1|J_+ |K>b^+_{3 K+1}b_{3 K}
\label{coriolis2_eq}
\end{eqnarray}
It was suggested by Bohr and Mottelson \cite{Vogel} to use the following
expression for the matrix element
$<K+1|J_+|K>$ in the case of the octupole excitations
\begin{eqnarray}
<K+1| J_+ |K>=\sqrt{(3-K)(3+K+1)}
\label{coriolis3_eq}
\end{eqnarray}
We will take, however,
\begin{eqnarray}
<K+1| J_+ |K>=\sqrt{(J_0 - K)(J_0 + K + 1)}
\label{coriolis4_eq}
\end{eqnarray}
considering $J_0$ as a free parameter. However, $J_0 \leq 3$.

Consider now a Coriolis coupling of the $K=0$ and $K=1$ states for
different angular momenta. In accordance with the arguments in
Section II the odd
$I$, $K=0$ states (\ref{coherent1_eq}) are shifted up in energy with 
respect to
the even $I$, $K=0$ states by an amount of energy $\widetilde{\Delta 
E(I)}$ (\ref{deltae_eq}).
We will assume that the $K$=1 phonon has an energy $\omega$.
In the nuclei considered below $\omega$ is much larger than 
$\widetilde{\Delta E}$,
however.

For every value of the angular momentum we have two states: one with 
$K$=0
(\ref{coherent1_eq}) and other with $K$=1 (\ref{coherent2_eq}). Their
energies are: 0 for $K$=0 and $\omega$+$\widetilde{\Delta E(I)}$ for 
$K$=1 and
even angular
momenta (positive parity), and $\widetilde{\Delta E(I)}$ for $K$=0 and 
$\omega$ for $K$=1
and odd angular momenta (negative parity).
Then,
considering a Coriolis interaction of the $K$=0 and $K$=1 states for 
different $I$ we have the following energy differences between mixing 
states:
$\omega +\widetilde{\Delta E(I)}$ if $I$ is even and
$\omega -\widetilde{\Delta E(I)}$ if $I$ is odd.
Thus, the energy
denominators for the Coriolis mixing are somewhat smaller
for odd angular momenta compared to the even angular momenta.
As a result the Coriolis coupling becomes somewhat more strong
for odd angular momenta
states. This effect is similar to that found for odd nuclei 
\cite{Leander,Brink} where
it explains a difference in the parity splitting in even $A$ and odd $A$
nuclei.

Consider now the matrix elements of the Coriolis interaction
(\ref{coriolis1_eq}) between the $K$=0 (\ref{coherent1_eq}) and 
$K$=1
(\ref{coherent2_eq}) states. The square of this matrix element is
proportional to an average number of the $b^+_{30}$ phonons in
the corresponding states. As it is discussed above in this section
the odd angular momenta states have a larger number of the $b^+_{30}$ 
phonons
than even angular momenta states. Therefore the Coriolis interaction
matrix elements are larger for odd angular momenta than for even
angular momenta states.
Together with the somewhat less important fact
that the energy differences between mixing states are somewhat
larger for the odd
$I$ states it means that the odd $I$ states of the alternating parity 
band
will be shifted down more than the even $I$ states. This effect can
explain the observed lowering of the odd I (negative--parity) states
relative to the even $I$ states, i.e. sign reversal of the parity 
splitting
at the large values of $I$ when the Coriolis interaction becomes strong.

When the octupole deformation is stabilized $\widetilde{\Delta E(I)}$
goes to zero, the
difference in the Coriolis matrix elements for odd and even $I$ states
goes to zero as shown in (\ref{split3_eq})
and we get an alternating parity band without staggering.

Let us introduce a quantity $\Delta E_{th}(I)$, which describes
parity splitting in the ground state alternating parity
band and, in contrast to $\widetilde{\Delta E(I)}$, includes both 
barrier penetration
and Coriolis mixing effects.
Using the standard formulae for the two--level mixing we get the 
following
expression for $\Delta E_{th}(I)$
\begin{eqnarray}
\Delta E_{th} (I)=\frac{\widetilde{\Delta E(I)}-\frac{1}{\omega}\left 
[F^2_- (I)-F^2_+ (I)\right 
]}{\sqrt{\frac{1}{4}(1+\frac{\widetilde{\Delta E(I)}}{\omega})^2 
+\frac{1}{\omega 
^2}F^2_+(I)}+\sqrt{\frac{1}{4}(1-\frac{\widetilde{\Delta 
E(I)}}{\omega})^2 +\frac{1}{\omega ^2}F^2_-(I)}}
\label{split1_eq}
\end{eqnarray}
where
\begin{eqnarray}
F^2_{\pm}(I)=\frac{1}{2}\left (\frac{\hbar ^2}{\Im}\right )^2 J_0 (J_0 
+ 1)I(I+1)n_{\pm}(I)
\label{split2_eq}
\end{eqnarray}
The quantity $\widetilde{\Delta E(I)}$ introduced in the preceding 
section
includes
only an effect of the barrier penetration.
Here $n_+ (I)$ ($n_- (I)$) is an average number of the $K=0$ octupole
bosons in the $I$ even (odd) states. Using the coherent states 
introduced
above and the Hamiltonian expressed in terms of the $b^+_{30}, b_{30}$ 
boson
operators describing axially symmetric octupole motion and having a 
potential
energy term with two octupole--deformed minima we can show numerically
that when octupole deformation becomes relatively stable the following
approximate expressions for $n_{\pm}(I)$ can be used
\begin{eqnarray}
n_+ (I)\simeq n \left (1-\left [\frac{\widetilde{\Delta 
E(I)}}{\widetilde{\Delta E(0)}}\right ]^{b} \right)\nonumber\\
n_- (I)\simeq \left [\frac{\widetilde{\Delta E(I)}}{\widetilde{\Delta 
E(0)}}\right ]^{b} + n \left (1-\left [\frac{\widetilde{\Delta 
E(I)}}{\widetilde{\Delta E(0)}}\right ]^{b} \right)
\label{split3_eq}
\end{eqnarray}
with $b=0.15\div 0.25$. The results shown below are obtained with 
$b$=0.15.
However, when $b$=0.25 that fit is not changed essentially.
In (\ref{split3_eq}) $n$ is a parameter depending on the
parameters of the Hamiltonian. We note that as follows from Eqs. 
(\ref{split1_eq}--\ref{split3_eq}) $\Delta E_{th}(I)\rightarrow 0$ if
$\widetilde{\Delta E(I)}\rightarrow 0$.

Now the expression (\ref{split1_eq}) for the parity splitting which
includes the effects
of the barrier penetration and of the Coriolis mixing with the $K=1$ 
octupole
vibrations is determined completely. It contains the following five
parameters:
$\widetilde{\Delta E(0)}$, $I_0$, $\frac{\hbar ^2}{\Im}\sqrt{J_0 (J_0 + 
1)}$, $\omega$ and $n$.
In the following section we will take two of these parameters, namely,
$\omega$ and $n$ to be fixed for all considered nuclei.
In some nuclei the value of $\omega$ is known.=20

\section{Description of the experimental data}

Since positive-- and negative--parity states have different angular =
momenta
(even and odd, correspondingly) parity splitting can be determined only
by interpolation of the experimental energies of the positive--parity 
states
to odd angular momentum and vice versa. Based only on the experimental
energies we have used the following formula to determine experimental
parity splitting
\begin{eqnarray}
\Delta E_{exp}(I)= (-1)^I\left (\frac{1}{2} 
[E_{exp}(I+1)-2E_{exp}(I)+E_{exp}(I-1) ]\right.\nonumber\\
\left.-\frac{1}{8}[E_{exp}(I+2)-2E_{exp}(I)+E_{exp}(I-2) ]\right)
\label{paritysplit_eq}
\end{eqnarray}
if $I\ge 2$, and
\begin{eqnarray}
\Delta E_{exp}(1)= -\left (\frac{1}{2} [E_{exp}(2)-2E_{exp}(1) 
]\right.\nonumber\\
\left.-\frac{1}{8}[E_{exp}(4)-2E_{exp}(3)]\right)
\label{paritysplit1_eq}
\end{eqnarray}
for $I$=1. We note that this formula differs slightly from the formula
for $\Delta E_{exp}(I)$ which was used in \cite{Jolos3}.
The quality of this formula can be verified by subtracting $\Delta 
E_{exp}(I)$
from the experimental energies $E_{exp}(I)$
\begin{eqnarray}
E_{av}(I)\equiv E_{exp}(I)+\frac{1}{2}(-1)^I \Delta E_{exp}(I)
\label{eaverage_eq}
\end{eqnarray}
The resulting function $E_{av}(I)$ should be a smooth function of $I$.
We have calculated $E_{av}(I)$ for all considered nuclei and found
that $E_{av}(I)$ is really a smooth function of $I$.

We consider the following nuclei: $^{220,222,224,226}$Ra and 
$^{222,224}$Th.
The data are taken from \cite{Cocks,Smith,Ackermann,Schuler,Schwartz}.
The experimental splitting $\Delta E_{exp}(I)$ is shown in Figs. (1-6).
It is compared to the theoretical splitting $\Delta E_{th}(I)$ obtained
from (\ref{split1_eq}--\ref{split3_eq}). In calculating
$\Delta E_{th}(I)$ we note that
the value of $\omega$ is known for $^{224}$Ra (1053 KeV) and
$^{226}$Ra (1048 KeV). Since these two energies are quite close we have
taken $\omega$=1050 KeV for all nuclei under consideration. We have 
also taken
$n$=3 for all considered nuclei to reduce the number of the free
parameters. This value is in a correspondence with the coherent
states parameters obtained in \cite{Vitturi}. Thus, to fit the 
experimental
spectra we use 3 parameters for every nucleus, namely, 
$\widetilde{\Delta E(0)}$,
$I_0$ and $\frac{\hbar ^2}{2\Im}\sqrt{J_0 (J_0 + 1)}$.

The results of the fit of the angular momentum dependence of the parity
splitting $\Delta E_{th}(I)$ are shown in Figs.1--6. The parameters used 
in the fit are presented
in Table 1, where for completeness the values of the attenuation factor
of the Coriolis interaction are presented.
If we estimate the value of $\hbar ^2$/$\Im$ fitting
$E_{av}(I)$ by the simple rotational formula
$(E_{av}(I+1)-E_{av}(I))=\frac{\hbar ^2}{\Im}(I+1)$ we get for all
considered nuclei the value of $J_0$ not exceeding 1.1. We get the 
minimum
value of $J_0$=0.6 for $^{226}$Ra. Thus, the values of the Coriolis 
strength
parameter needed to fit the spectra are quite reasonable.
The corresponding attenuation factor of the Coriolis interaction takes
the values between 0.28 and 0.44. We have performed also the 
calculations with
the same attenuation factor for all considered nuclei, which we have 
taken
to be equal to 0.33. In this case the results for parity splitting in
$^{222,224}$Ra and $^{224}$Th are practically coincide with those shown
in Figs.2,3 and 5. The results for $^{226}$Ra and $^{222}$Th deviate 
slightly
from the preceding ones. However, these deviations do not exceed 6 KeV
for $^{226}$Ra and 5 KeV for $^{222}$Th. The results obtained for 
$^{220}$Ra
with common attenuation factor deviate from the best fit obtained for
this nucleus more than for
other considered nuclei. Namely, these deviations take the values up
to 20 KeV. The results obtained with common attenuation factor 0.33 for
$^{220,226}$Ra are presented in Figs.1 and 4.

The value of $\widetilde{\Delta E(0)}$ characterize a stability of the 
octupole
deformation in the ground state. This quantity is connected with the 
freequency
of transition through the octupole barrier. The smaller the value of
$\widetilde{\Delta E(0)}$, the smaller the value of the transition
freequency, the larger the barrier height separating two symmetrically
located octupole minima with different signs of the octupole 
deformation.
For nuclei considered in this paper $\widetilde{\Delta E(0)}$ takes the
values between 200 KeV and 300 KeV. These values are the smallest ones
among those known for the observed alternating parity bands. In Ra 
isotopes
the value of $\widetilde{\Delta E(0)}$ decreases initially when we go 
from
$^{226}$Ra to lighter isotopes, indicating an increase of the strength
of the octupole correlations. However, it increases again in $^{220}$Ra.
Probably, it is connected with the fact that $^{220}$Ra is a 
transitional
nucleus with $E(2^+_1)$=178 KeV. In $^{226}$Ra $E(2^+_1)$=68 KeV. 
The same
effect is observed in Th isotopes.
As is seen
the largest
deviations of the fit from the experimental data are obtained for 
$^{220}$Ra and
$^{222}$Th. These two nuclei are rather transitional ones, however.
\cite{Cocks}.

\section{Conclusion}

A model is developed to describe the angular momentum dependence of the
parity splitting in the nuclei with strong octupole correlations.
The model explains the positive parity splitting at the relatively low 
angular
momenta and its dependence on $I$ as produced by a penetration of the 
barrier
separating two symmetrically located octupole minima having different
signs of the octupole deformation. It is demonstrated that the sign
reversal of the parity splitting at higher values of $I$ can be 
interpreted
as a result of the Coriolis mixing with $K$=1 octupole mode. However,
a consideration of this effect is qualitative and a more detailed 
treatment
requires a microscopic approach. In some nuclei parity splitting changes
the sign in second time \cite{Mikhailov} although its absolute value
decreases in average with $I$ increase. Probably, it means that the
Coriolis mixing among several intrinsic states should be considered.

\acknowledgments
The authors would like to express their gratitude  to Dr.R.--D. Herzberg
for reading the manuscript. The work was supported in part by the BMFT
under contract 060 K 602 I.
One of the authors (R. V. J) is grateful to the
Universit\"at zu K\"oln, where this work has been done, for its 
support.

\begin{figure}
\centerline{\epsfig{file=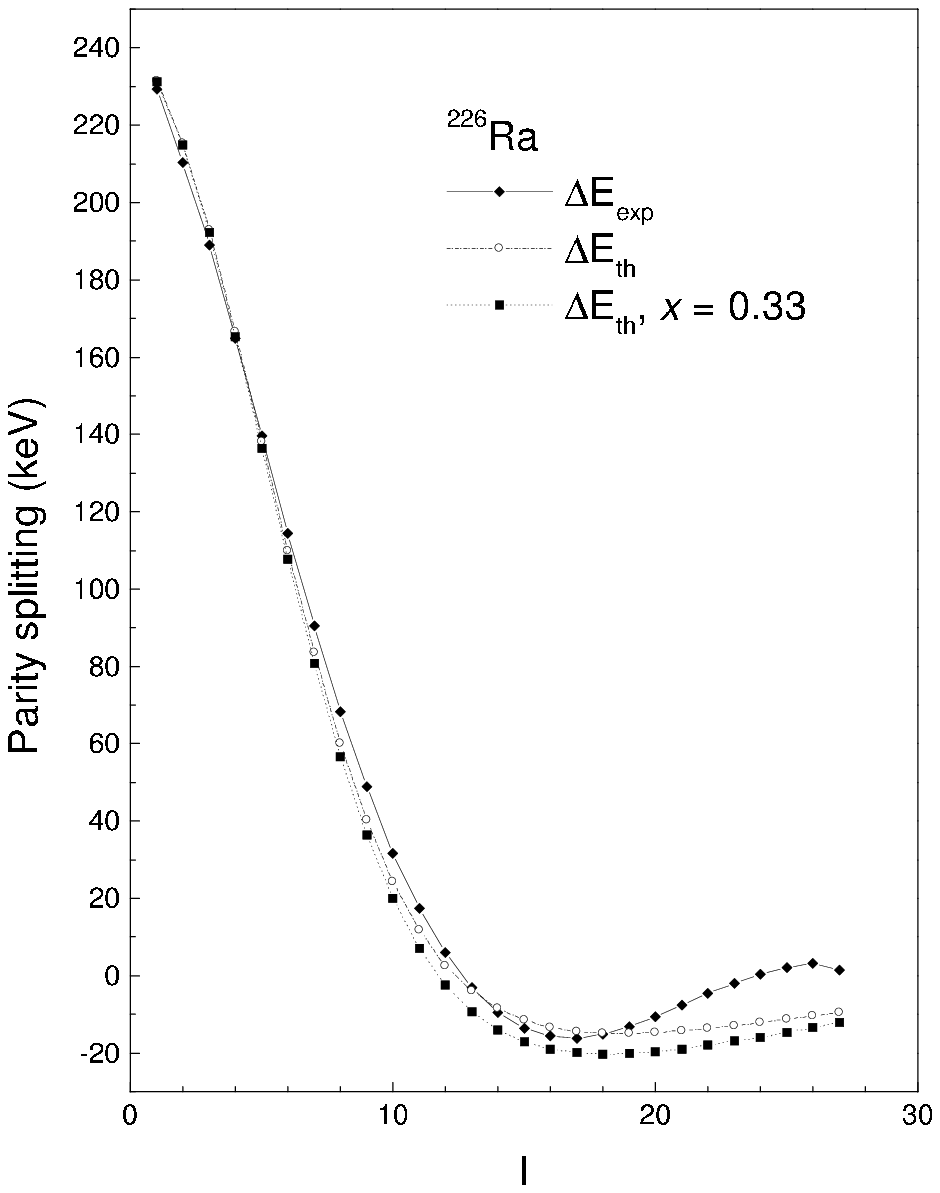,width=6in}}
\vspace*{10pt}
\caption{Comparison of the experimental $\Delta E_{exp}(I)$ and 
calculated
$\Delta E_{th}(I)$ values of the parity
splitting for $^{226}$Ra. The results of
calculations obtained with the common attenuation factor for all
considered nuclei ($\Delta E_{th}$, $x$=0.33) are shown in addition.
The fit parameters are given
in Table 1. The experimental data are taken from \protect\cite{Cocks}}
\label{ra226_fig}
\end{figure}

\begin{figure}
\centerline{\epsfig{file=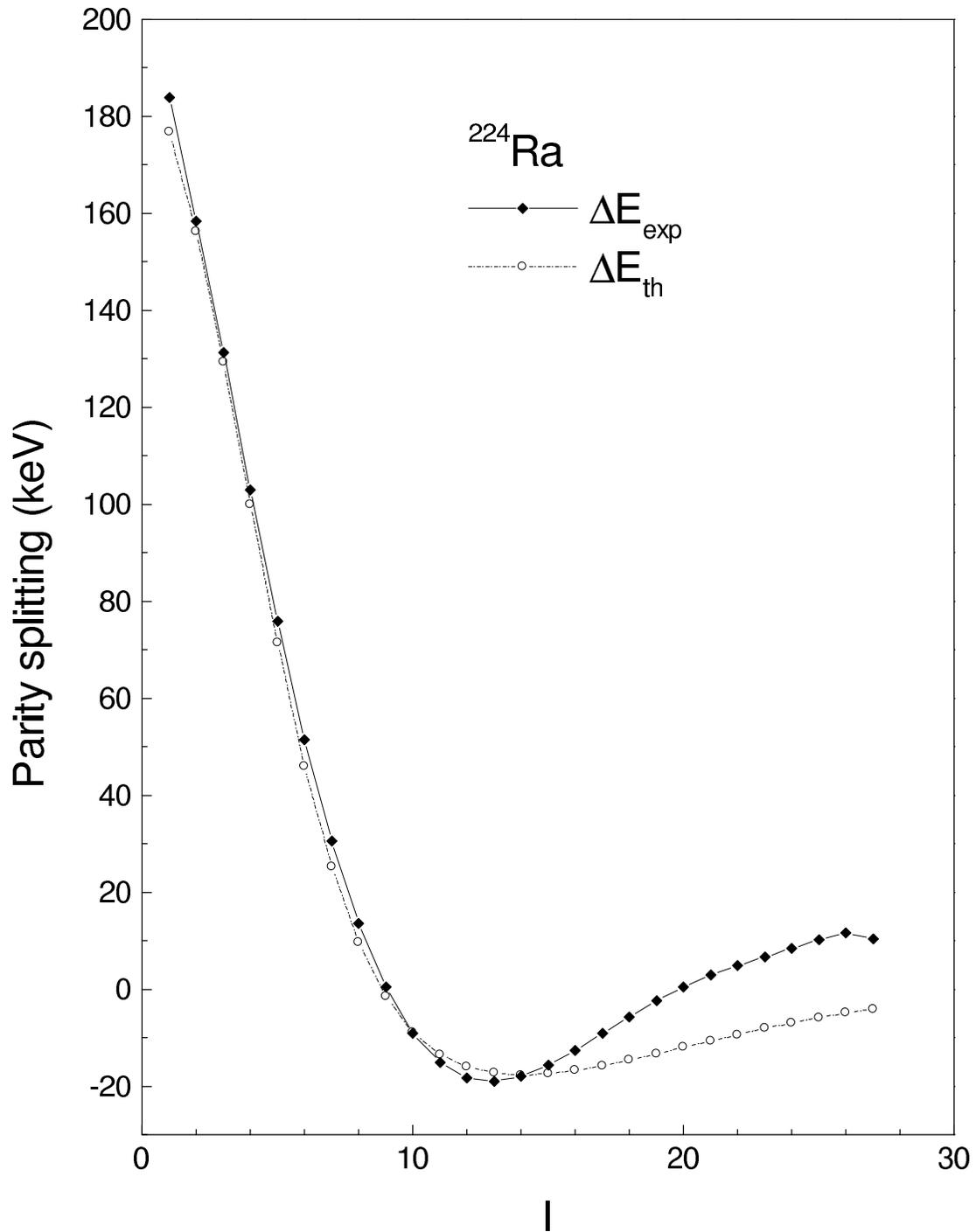,width=6in}}
\vspace*{10pt}
\caption{Comparison of the experimental $\Delta E_{exp}(I)$ and 
calculated
$\Delta E_{th}(I)$ values of the parity splitting for $^{224}$Ra.
The fit parameters are given in Table 1. The experimental data
are taken from \protect\cite{Cocks} }
\label{ra224_fig}
\end{figure}

\begin{figure}
\centerline{\epsfig{file=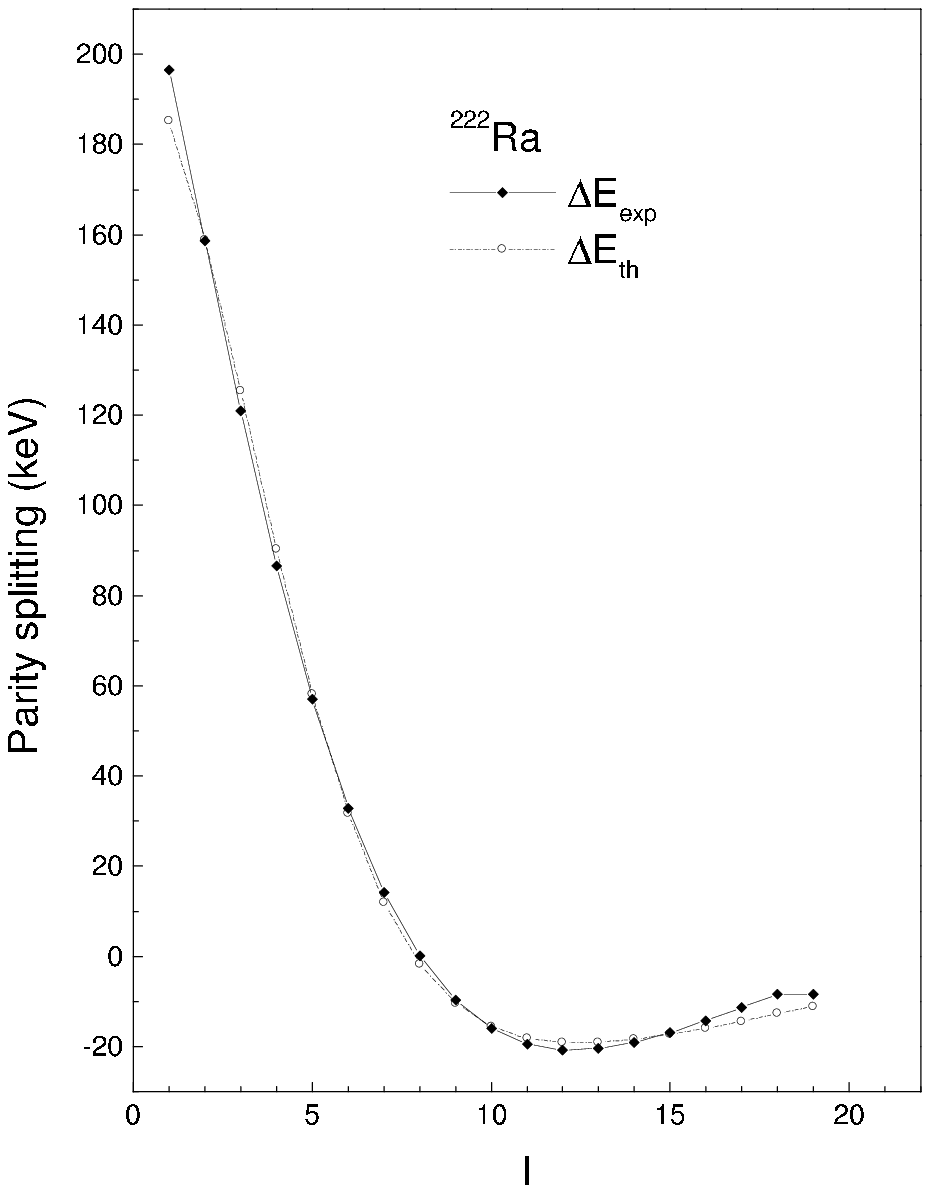,width=6in}}
\vspace*{10pt}
\caption{The same as in Fig.2 but for $^{222}$Ra. The experimental data
are taken from \protect\cite{Cocks} }
\label{ra222_fig}
\end{figure}

\begin{figure}
\centerline{\epsfig{file=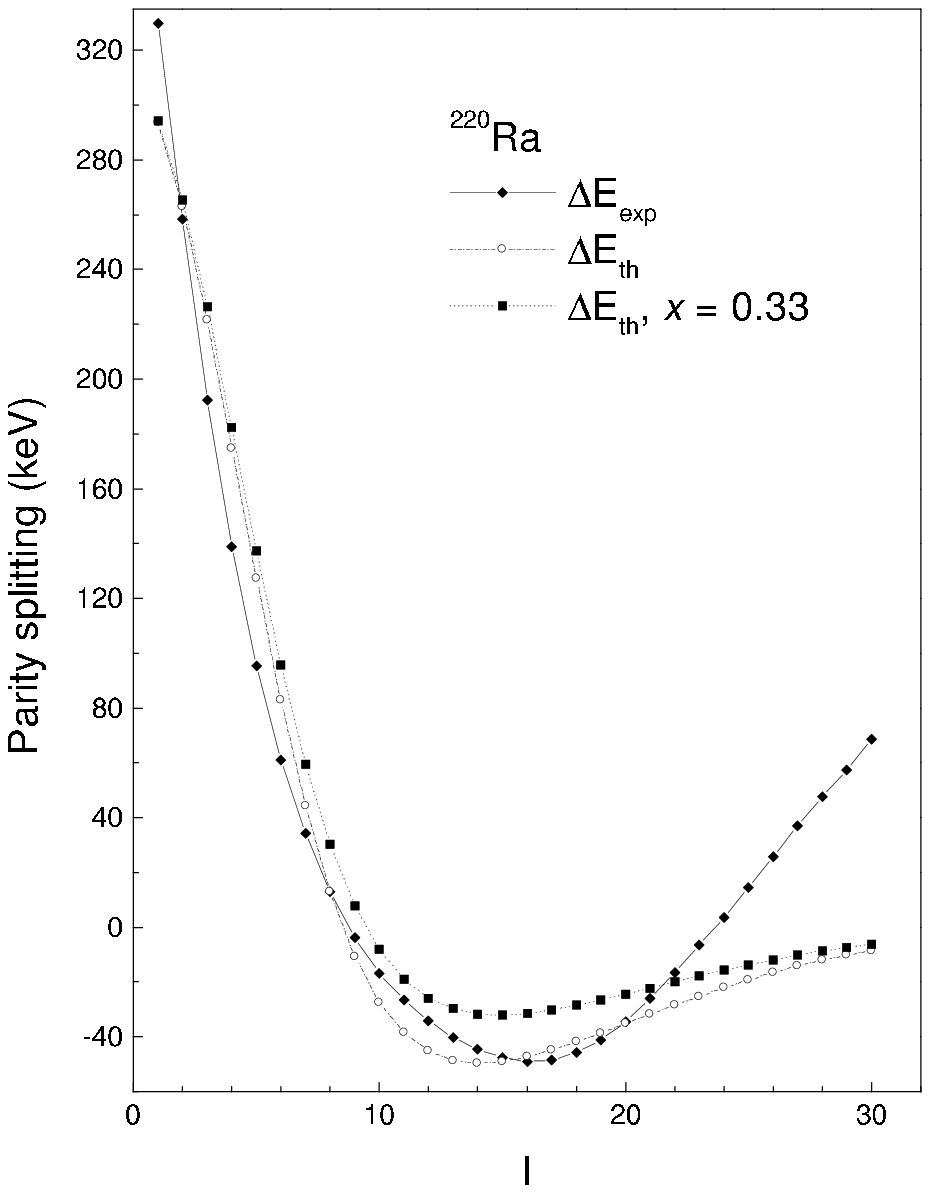,width=6in}}
\vspace*{10pt}
\caption{The same as in Fig.1 but for $^{220}$Ra. The experimental data
are taken from \protect\cite{Smith} }
\label{ra220_fig}
\end{figure}

\begin{figure}
\centerline{\epsfig{file=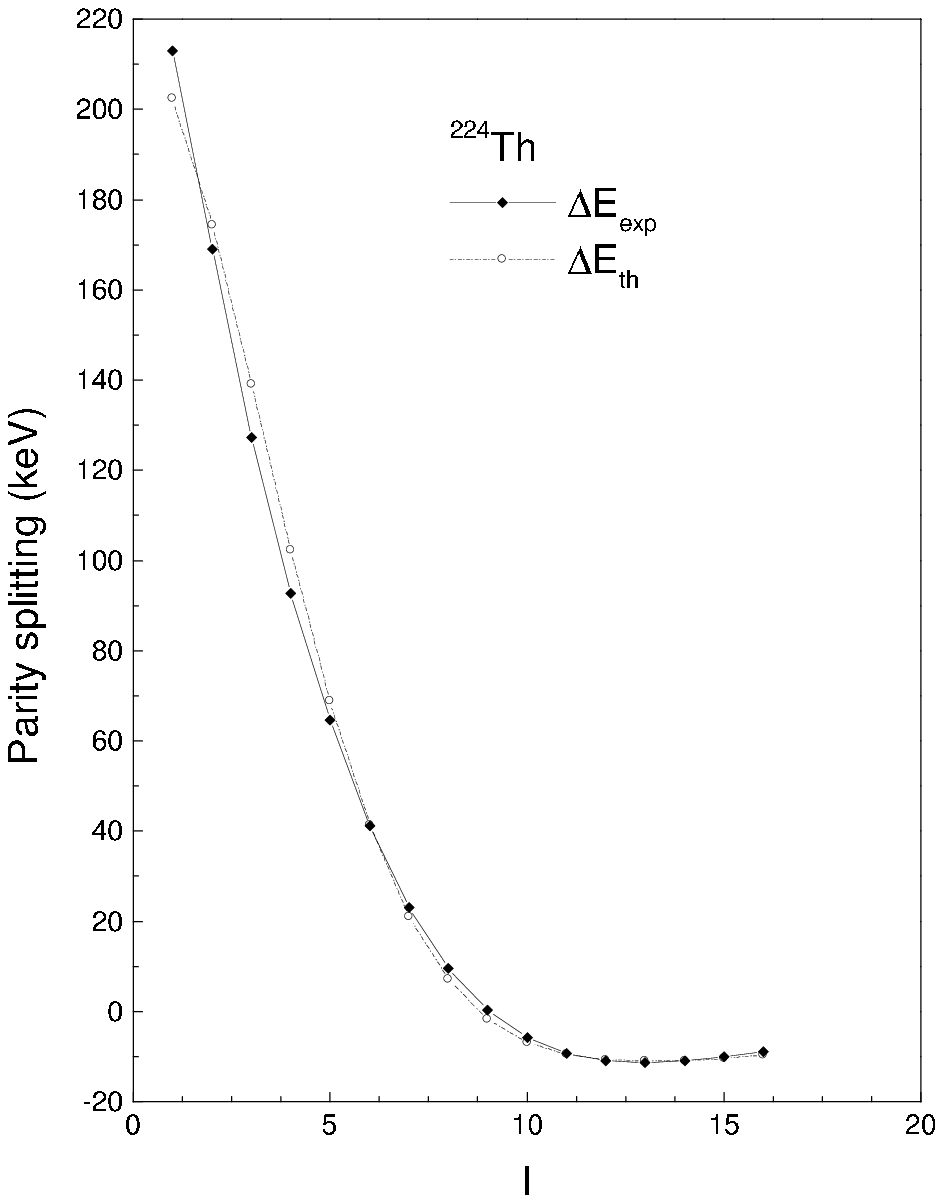,width=6in}}
\vspace*{10pt}
\caption{The same as in Fig.2 but for $^{224}$Th. The experimental data
are taken from \protect\cite{Ackermann,Schuler} }
\label{th224_fig}
\end{figure}

\begin{figure}
\centerline{\epsfig{file=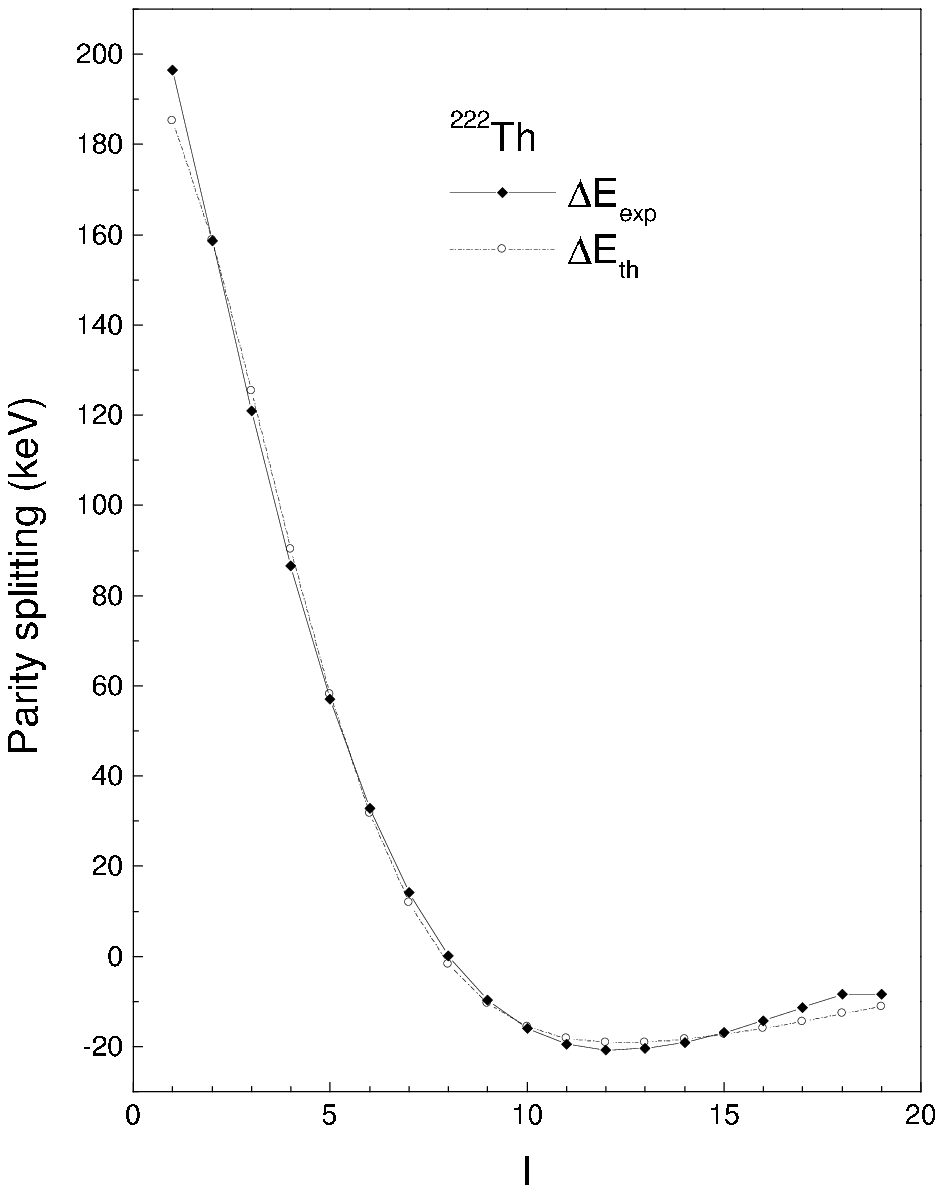,width=6in}}
\vspace*{10pt}
\caption{The same as in Fig.2 but for $^{222}$Th. The experimental data
are taken from \protect\cite{Cocks,Schwartz}  }
\label{th222_fig}
\end{figure}

\begin{table}
\caption
{Values of the parameters used for fit of $\Delta E_{exp}(I)$ using
Eqs.(\protect\ref{split1_eq}--\protect\ref{split3_eq}) and $n$=3, 
$\omega$=1050 KeV.
The values of the attenuation factor of the Coriolis interaction $x$ are
shown for completeness. }
\label{1_tab}
\begin{tabular}{cddddd}
   {\bf Nucleus} & Ref. & $\widetilde{\Delta E(0)}$[KeV] & $I_{0}$ & 
$\frac{\hbar ^2}{2\Im}\sqrt{J_0 (J_0 + 1)}$[KeV]  & $x$\\
   \tableline
   $^{226}$Ra & [3] & 240 & 7.1 & 16.3 & 0.28   \\ 
   $^{224}$Ra & [3] & 188 & 5.4 & 23.1 & 0.33 \\
   $^{222}$Ra & [3] & 200 & 4.8 & 27.0 & 0.34  \\
   $^{220}$Ra & [18] & 310 & 6.0 & 39.6 & 0.44  \\
   $^{224}$Th & [19,20] & 218 & 4.8 & 19.8 & 0.30 \\
   $^{222}$Th & [3,21] & 305 & 4.6 & 36.4 & 0.37  \\
\end{tabular}
\end{table}

\end{document}